\documentclass[final,5p,times,twocolumn,authoryear]{elsarticle}

\usepackage{amsmath,amsfonts,amsthm,amssymb}
\usepackage{stfloats}
\usepackage{glossaries}
\usepackage{placeins}
\usepackage{subcaption}
\usepackage{xcolor}
\usepackage{balance}

\def\Rbb{\mathbb{R}}

\def\Bcal{\mathcal{B}}

\def\Fcal{\mathcal{F}}

\def\pdot{\dot{p}}
\def\vdot{\dot{v}}
\def\Rdot{\dot{R}}
\def\thetadot{\dot{\theta}}
\def\omegadot{\dot{\omega}}
\def\pddot{\ddot{p}}
\def\omegahat{\hat{\omega}}

\def\thetaddot{\ddot{\theta}}
\def\mbar{\bar{m}}
\def\na{{n_a}}

\def\ehat{\hat{e}}

\newcommand{\inner}[1]{\langle #1 \rangle}
\newcommand{\col}[1]{\begin{bmatrix}#1\end{bmatrix}}

\newtheorem{remark}{Remark}

\newacronym{LOE}{LOE}{Loss Of Effectiveness}
\newacronym{FD}{FD}{Fault Detection}
\newacronym{NED}{NED}{North-East-Down}
\newacronym{ESC}{ESC}{Electronic Speed Controller}
\newacronym{CCW}{CCW}{CounterClockWise}
\newacronym{CW}{CW}{ClockWise}
\newacronym{FDI}{FDI}{Fault Detection and Isolation}
\newacronym{FI}{FI}{Fault Isolation}
\newacronym{DFT}{DFT}{Discrete Fourier Transform}
\newacronym{AFD}{AFD}{Active Fault Diagnosis}
\newacronym{UAV}{UAV}{Unmanned Aerial Vehicle}
\newacronym{COM}{COM}{Center Of Mass}
\newacronym{IMU}{IMU}{Inertial Measurement Unit}
\newacronym{SVM}{SVM}{Support Vector Machine}
\newacronym{FDD}{FDD}{Fault Detection and Diagnosis}
\newacronym{FEA}{FEA}{Finite Element Analysis}
\newacronym{LSTM}{LSTM}{Long and Short-Term Memory}
\newacronym{CL}{CL}{Continual Learning}
\newacronym{FFT}{FFT}{Fast Fourier Transform}
\newacronym{DNN}{DNN}{Deep Neural Network}
\newacronym{FTC}{FTC}{Fault Tolerant Control}
\newacronym{SVMM}{SVMM}{Spatial-Vibration-Modulation-Assisted}
\newacronym{ANN}{ANN}{Artificial Neural Network}
\newacronym{TPR}{TPR}{True Positive Rate}
\newacronym{FPR}{FPR}{False Positive Rate}

\journal{ArXiv}

\begin{document}
\begin{frontmatter}

\title{Active Propeller Fault Detection and Isolation in Multirotors Via Vibration Model}

\author[label1]{Alessandro Baldini}
\author[label1]{Riccardo Felicetti}
\author[label1]{Alessandro Freddi}
\author[label1]{Andrea Monteriù}
\affiliation[label1]{organization={Department of Information Engineering, Università Politecnica delle Marche},
            addressline={Via Brecce Bianche 12},
            city={Ancona},
            postcode={60131},
            country={Italy}}

\begin{abstract}
    In rotary-wing aircraft, rotating blades are exposed to collisions and subsequent damage.
    The detection and isolation of blade damage constitute the first step in fault mitigation; however, they are particularly challenging when considerable input redundancy is available, as in the case of multirotors.
    In this article, we propose an active model-based approach that deliberately perturbs the control inputs to isolate blade faults in multirotor vehicles.
    By exploiting a model that captures the vibrations caused by blade damage, the isolation method relies solely on vibration data from the onboard inertial measurement unit.
    The strategy is tested in simulation using an octarotor platform, and both time-domain and frequency-domain features are analyzed.
    Several accuracy-related metrics of the technique are evaluated on a set of $9600$ simulations and compared with the most relevant variables.
\end{abstract}




\end{frontmatter}

\section{Introduction}

\glspl{UAV} have become integral to a wide range of military and civilian activities, offering cost-effective capabilities for applications such as surveillance, search and rescue, infrastructure inspection, logistics, aerial imaging, and intelligence gathering.
The increasing diffusion of \glspl{UAV} in commercial and private contexts, together with their widespread use by non-professional and recreational users, has intensified concerns related to safety, privacy, and security \citep{tan2021public}.
These concerns are particularly relevant for small-scale platforms, which exhibit comparatively high failure rates \citep{shraim2018survey}, largely attributable to equipment malfunctions.
Empirical evidence indicates that equipment failures account for a significant proportion of incidents, with approximately 64\% of accidents involving remotely piloted \glspl{UAV} being linked to such causes \citep{wild2016exploring}.
Among these, actuator-related faults are especially critical, as they may rapidly compromise flight stability, reduce control authority, and lead to vehicle damage if not promptly addressed.
In this context, \gls{FD} represents a fundamental component of fault-tolerant operation, serving as the initial step in fault mitigation.
To be effective across diverse platforms and operating conditions, \gls{FD} schemes should therefore be designed with a high degree of robustness and adaptability to different environmental conditions and \gls{UAV} configurations \citep{10387779}.

A substantial body of literature investigates the use of vibration signals and frequency-domain features for the detection of propeller faults in \glspl{UAV}.
Acceleration-based approaches are explored in \cite{ghalamchi2018vibration}, where propeller damage is identified through onboard measurements; however, the proposed method is limited to scenarios in which the vehicle follows a predefined trajectory.
An extension is presented in \cite{ghalamchi2019real}, where an Extended Kalman Filter is employed to estimate mass unbalance induced by a chipped blade, demonstrating the feasibility of fault isolation.
Despite its effectiveness, the reported convergence time for a medium-sized hexarotor is on the order of several minutes, which may be unsuitable for safety-critical applications.
More recent contributions focus on real-time implementations, such as the vibration-based \gls{FD} strategy proposed in \cite{baldini2023real}.
In \cite{rao2025real}, experimental validation of stochastic subspace-based \gls{FD} methods combined with histogram-based analysis of acceleration data is provided for both hovering and linear flight conditions.

Most existing studies primarily focus on \gls{FD}, as isolating the specific faulty propeller during flight remains a challenging task.
This difficulty arises from the large number of vibration-related frequency components that accumulate during operation, making individual sources of damage hard to distinguish.
As a result, approaches that explicitly address propeller \gls{FDI} and \gls{FDD} tend to rely on data-driven methods with relatively high model complexity.
A representative example is provided in \cite{2017_SPA}, where a \gls{FDD} framework based on \gls{IMU} acceleration measurements is introduced.
Flight data collected under different propeller damage conditions are used to extract time- and frequency-domain features, which are subsequently employed to train a one-vs-all \gls{SVM} classifier.
The same authors extend this analysis in \cite{9874283}, deploying a lightweight \gls{ANN} on embedded hardware for online fault classification.
Similar data-driven strategies are adopted in later works, such as \cite{zhang2021fault}, where time--frequency features are processed by an \gls{LSTM} network for propeller \gls{FDI}.
In \cite{s22166037}, vibration analysis is further expanded to include gyroscopic data acquired through a custom hardware setup and clustered using a K-means algorithm.
Recently, multimodal approaches combining accelerometer, gyroscope, and control energy information have been proposed, with a \gls{DNN} directly regressing the extent of propeller damage in millimeters \cite{10156355}, later extended to integrated damage estimation in \cite{10912747}.
In parallel, several public datasets have become available and are widely used to benchmark data-driven \gls{FDI} and \gls{FDD} methods.
Notable examples include the dataset presented in \cite{baldini2023uav}, which provides inertial and flight controller data, and the dataset in \cite{puchalski2024padre}, which combines inertial measurement unit data with microphone recordings.

Although the aforementioned approaches are able to achieve \gls{FDI} and \gls{FDD}, they generally lack a formal justification of the vibration frequencies expected under specific operating and flight conditions.
In the absence of an explicit vibration model, data-driven \gls{FDI} and \gls{FDD} techniques tend to be highly sensitive to changes in flight maneuvers, payload, and environmental conditions.
As a consequence, these methods often require continual retraining as new data become available, which hampers reproducibility and limits their applicability when the experimental setup or vehicle configuration is modified.
To address these limitations, this work adopts an \gls{AFD} framework grounded in an explicit vibration model.

\gls{AFD} improves fault detection performance by deliberately injecting a designed auxiliary input into the system and analyzing the resulting input--output response.
This paradigm alleviates several shortcomings of passive approaches, particularly in the presence of model uncertainty, noise, disturbances, or faults whose effects are weak or partially masked \cite{heirung2019input}.
When combined with vibration analysis, \gls{AFD} becomes especially relevant, as variations in lift and drag induced by blade chipping may be indistinguishable from inherent motor performance variability \cite{ghalamchi2019real}.
Moreover, for overactuated \glspl{UAV}, the design of auxiliary control inputs that minimally affect nominal dynamics can be carried out in a systematic and straightforward manner.

From a modeling perspective, vibration analysis in the literature is often carried out using high-fidelity tools such as \gls{FEA} \cite{2019_IOP} or computational fluid dynamics \cite{jiang2025damage}.
While these techniques provide detailed insights into structural and aerodynamic effects, their computational complexity renders them unsuitable for online \gls{FDI}.
Simpler modeling approaches, such as those proposed in \cite{zou2025dynamics}, are better suited to capture the dominant vibration phenomena associated with blade mass unbalance.
In this work, a rigid-body formulation is adopted to derive the effects of blade damage on the vehicle's linear and angular accelerations, complemented by blade element theory \cite{bangura2016aerodynamics}.
The resulting model captures the essential vibration-induced dynamics with sufficient fidelity, while remaining computationally tractable for online \gls{FDI} and \gls{FTC} applications in \glspl{UAV}.

Using a multi-rigid model that captures the oscillatory dynamics of the chipped propellers, in this paper we design a model-based active \gls{FDI} strategy to isolate the faulty blade.
The technique intentionally alters the control input to isolate the faulty propeller.
The provided solution makes use of the well known frequency domain features, whose policies are guided by the geometry and the dynamics of the system.
The technique is discussed for multirotors having coplanar and collinear propellers, which represents the most common configuration, and for which the linearly dependent trust pointing vectors are make the isolation challenging.
However, the similar approach can be extended for variable tilting and morphing configurations.

The paper is organized as follows.
In Section~\ref{sec:model}, we present the multirotor model, which accounts for the rigid-body representation of blade chipping.
Section~\ref{sec:frequency} analyzes the vibration from a frequency-domain perspective, detailing low- and high-frequency components.
Based on the these considerations, Section~\ref{sec:FDI} presents the proposed active method to isolate faults due to blade chipping.
Extensive simulation results are shown in Section~\ref{sec:simresults} to validate the method across a wide range of conditions, and concluding remarks are provided in Section~\ref{sec:conclusion}.

\section{Multirotor model}\label{sec:model}
\begin{figure}[t]
	\centering
	\includegraphics[width=0.7\linewidth]{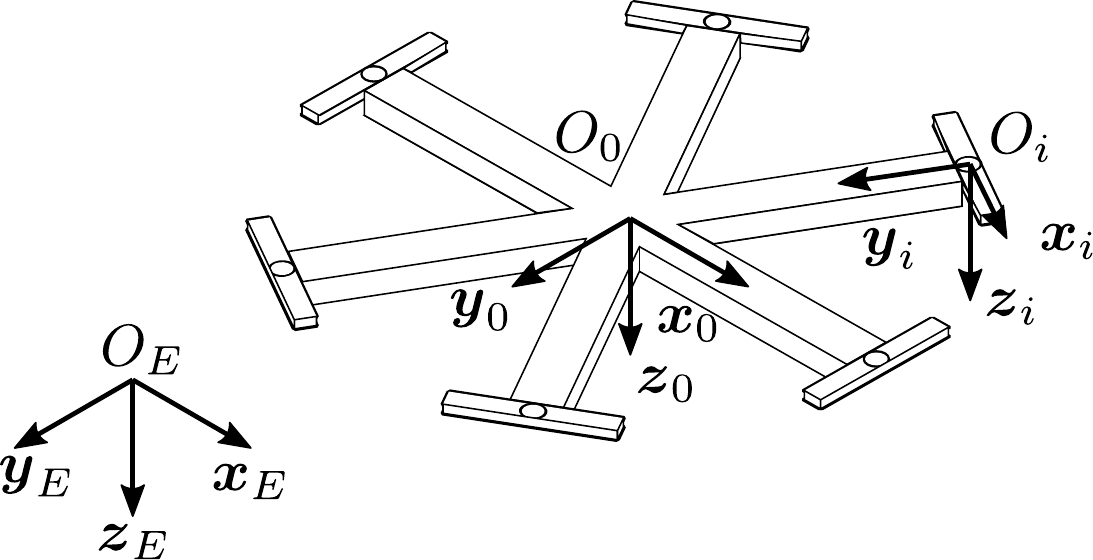}
	\caption{NED reference frames for a standard hexarotor.}
	\label{fig:frames1}
\end{figure}

In this paper we consider a multirotor with $n_a$ coplanar and collinear propellers, which is modeled as a constrained system of $n_a+1$ rigid bodies, i.e., one for each rotating propeller and the remaining one for the central main body on which the \gls{IMU} and the propellers are attached.
More precisely, let $R_E = (O_E, x_E, y_E, z_E)$ be a \gls{NED} earth-fixed reference frame, assumed to be inertial, let $R_0 = (O_0, x_0, y_0, z_0)$ be a body-fixed reference frame, centered in the center of mass of the central body, and let $R_i = (O_i, x_i, y_i, z_i)$ be a body-fixed reference frame in the center of mass of the $i$th propeller, for $i=1,\ldots,n_a$.
An example as depicted in Fig.~\ref{fig:frames1} for a standard hexarotor having $n_a =6$ actuators.
Without loss of generality, we assume the positions of the \gls{IMU} and $R_0$ coincide.
Therefore, we consider the generic \gls{UAV} described by the mathematical model
\begin{subequations}\label{eq:motion}
    \begin{align}
        &\dot{p}_{E,0}^E 
        = R_0^E v_{E,0}^0\>,\label{eq:multirotor_linkinematics}\\
    	&\Rdot_0^E = R_0^E \omegahat_{E,0}^0\>,\label{eq:multirotor_angkinematics}\\
    	&m(\vdot_{E,0}^0+\omega_{E,0}^0\times v_{E,0}^0) +  f_{as}^0 + f_{aa}^0 
        = 
        f_{g}^0 + f_m^0 + f_{fr}^0\>,\label{eq:multirotor_lindynamics}\\
    	&I\dot{\omega}_{E,0}^0 + {\omega}_{E,0}^0\times I{\omega}_{E,0}^0 + \tau_{gyro}^0
        + \tau_a^0
        = 
        r_{0,g}^0 \times f_g^0 + \tau_{fr}^0 + \tau_m^0\>,\label{eq:multirotor_angdynamics}\\
    	&\ddot{p}_i 
        = 
        \thetaddot_i (e_3\times p_i) + \thetadot_i(e_3 \times \dot{p}_i) \qquad i=1,\ldots,\na\>,\label{eq:multirotor_propeller}
    \end{align}
\end{subequations}
where $p_{E,0}^E\in\Rbb^3$ is the position of $R_0$ relative to $R_E$ (expressed in $R_E$), $v_{E,0}^0\in\Rbb^3$ denotes the velocity of $R_0$ relative to $R_E$ (expressed in $R_0$), $R_0^E\in SO(3)$ is the rotation matrix mapping $R_0$ to $R_E$, and $\omega\in\Rbb^3$ is the angular velocity of $R_0$ relative to $R_E$ (expressed in $R_0$).
Finally, $p_i = col(\cos(\theta_i),\sin(\theta_i),0)$ denotes the configuration of the $i$th blade, i.e., it models the pure rotation of angle $\theta_i$ around a fixed and vertical rotation axis in $R_0$, while $e_3=col(0,0,1)$.
It is important to emphasize that the parameters, the forces, and the torques describing model \eqref{eq:motion} are affected by possible blades chippings.
Thus, in the remainder of the section we briefly summarize the most relevant terms.

\subsection{Mass, center of mass and inertia}
The propeller of the $i$th actuator is modeled as a homogeneous rigid rod having two blades with possibly different radii $r_{1i},r_{i2}\in [0,r]$, where $r>0$ denotes the length of each healthy blade.
When the $i$th blade suffers from chipping, $r_{1i}$ and/or $r_{i2}$ differ from $r$.
For $i=1,\ldots,n_a$, the quantities 
\begin{align}
	w_{i} &= \frac{1}{2r}(r_{i1}+r_{i2})&
	w_{k_i} &= \frac{1}{r}(r_{i1}-r_{i2})
\end{align}
take into consideration the magnitude and the asymmetry of the chipping of the $i$th propeller, respectively.
It follows that the multirotor total mass is $m = m_0 + \sum_{i=1}^{n_a} w_i\mbar_i$, where $m_0$ denotes the mass of the central body, while $\mbar_i$ denote the mass of the $i$th propeller in the nominal scenario, i.e., when $r_{i1}=r_{i2}=r$.
Consequently, the center of mass of the multirotor (expressed in $R_0$), which is denoted with $r_{0,g}^0$, is $r_{0,g}^0 = r_{g,l}^0 + r_{g,h}^0$, where
    \begin{align}
    	r_{g,l}^0 &= \frac{1}{m} \sum_{i=1}^{n_a} m_il_i^0
    	&
    	r_{g,h}^0 &= \frac{1}{m} \sum_{i=1}^{n_a} m_i k_ip_i,
    \end{align}
$k_i = \frac{r}{2}w_{k_i}$, and $l_i^0$ denotes the vector from $R_0$ to the $i$th propeller attaching point (expressed in $R_0$).
Finally, the total time varying inertia tensor of the multirotor (expressed in $R_0$) is $I = I_0 + \sum_{i=1}^\na R_i^0 I_i (R_i^0)^T$, where $I_0$ is the inertia tensor of the central body (expressed in $R_0$), and $I_i$ is the inertia tensor of the $i$th blade w.r.t. its rest frame $R_{i}$ (expressed in $R_i$).
Please note that $I_i$ is affected by the blade chipping.

\subsection{External and and actuation wrenches}
The main external disturbances to take into account are those provided by the gravity and the air drag.
Therefore, $f_g^0 = mg(R_0^E)^Te_3$ denotes the force due to the gravity, while the linear and angular frictions are $f_{fr}^0 = -k_t (R_0^E)^Tv_{E,0}^0$, and $\tau_{fr}^0 = -k_r \omega_{E,0}^0$, where $k_t,k_r>0$ denote the friction coefficients \citep{baldini2020actuator}.

The pair $(f_m^0,\tau_m^0)$ denotes the wrench which can be used for control purposes (expressed in $R_0$), which is affected by the blade faults as well.
More precisely, using $c_{w_i}\in\{0,1\}$ to denote the clockwise/counterclockwise rotation of the $i$th propeller, the actuation force $f_m^0$ is $f_m^0 = \sum_{i=1}^\na f_{m,i}^0$, where 
\begin{align}
    f_{m,i}^0  
    &= 
    w_i c_{w_i}c_L \thetadot_i |\thetadot_i| e_3.
\end{align}
The actuation torque $\tau_m^0$ represents the overall torques which directly depends from the propeller rotation, i.e., from the upward lifts, and it can be calculated as
\begin{equation}
    \tau_{m}^0 
    = 
    \sum_{i=1}^\na (l_i^0\times f_{m,i}^0) + \sum_{i=1}^\na (\tau_{d,i}^0+\tau_{a,i}^0),
\end{equation}
where $\tau^0_{d,i} = -w_{d,i} c_D \thetadot_i |\thetadot_i| e_3$, $\tau^0_{a,i} = w_{a,i} c_D \frac{2}{3r}\thetadot_i |\thetadot_i|e_3$, and
\begin{align}
	w_{d,i} &= \frac{r_{i,1}^4+r_{i,2}^4}{2r^4},
    &
	w_{a,i} &= \frac{(r_{i,1}-r_{i,2})(r_{i,1}^3-r_{i,2}^3)}{2r^4}.
\end{align}
Finally, $c_D = \frac{1}{2}k_dr^4$ denotes the nominal drag coefficient, where $k_d$ is a constant.

\subsection{Vibration features and damping}
The blade chippings provide small variations of the center of mass and the mass.
Coupled with the high frequency rotation of the propellers, they generate additional forces and torques.
Two additional forces, denoted with $f_{as}^0$ and $f_{aa}^0$, and expressed as
\begin{align}
	f_{as}^0 
	&= 
	\sum_{i=1}^\na m_i(\omega_{E,0}^0 \times (\omega_{E,0}^0 \times l_i^0) + \omegadot_{E,0}^0 \times l_i^0)\\
	f_{aa}^0
	&= 
	\sum_{i=1}^\na                m_ik_i(\omega_{E,0}^0\times(\omega_{E,0}^0 \times p_i)) \\
    &+ \sum_{i=1}^\na 
    m_ik_i( 2 (\omega_{E,0}^0\times \pdot_i) 
	+ \dot{\omega}_{E,0}^0 \times p_i + \pddot_i),\nonumber
\end{align}
are injected into the plant due to the blade chipping.
Please note that, in absence of chipping, we consider a symmetric structure, and hence $f_{as}^0=0$.
Thus, due to the conservation of the angular momentum, the torque $\tau_a^0$ and the gyroscopic torque $\tau_{gyro}^0$ are expressed as
\begin{align}
    \tau_a^0 
    &= \sum_{i=1}^\na (p_{0,i}^0 \times (m\vdot_{E,0}^0+m(\omega_{E,0}^0\times v_{E,0}^0) + f_{as}^0 + f_{aa}^0))\\
    \tau_{gyro}^0
	&= \sum_{i=1}^\na 
	\left(
    \thetadot_i [\ehat_3,I_i^0]\omega_{E,0}^0 + I_{iz}\thetadot_i(\omega_{E,0}^0\times e_3)+ I_{iz}\thetaddot_i e_3\right),
\end{align}
where $I_{iz}$ is the component $(3,3)$ of $I_i$, consequently modifying the wrench acting on the plant.

Despite the fact the such vibrational contributions can be captured by the \gls{IMU}, the structure is not actually rigid.
Thus, we need to consider harmonic distortion and magnitude damping due to structure elasticity.
The first issue has been examined in the literature, showing that multiples of the fundamental frequency are present, with the lower-order harmonics exhibiting the greatest significance \citep{balandi2025enhancing,baldini2023real}.
The second source of damping arises from the structure itself and from the \gls{IMU}, which is not rigidly attached to the \gls{UAV}. This damping phenomenon is commonly modeled as a mass–spring–damper system, i.e., a linear system \citep{balandi2025enhancing}.
Consequently, for a given rotational speed, the cumulative damping effect results in an attenuation that must be applied to the vibration signals modeled in the following sections.

\section{Vibrations and loss of effectiveness}\label{sec:frequency}
In this section we study the forces (and thus the accelerations/vibrations) back to the \gls{IMU} provided by the blade fault due to chipping.
Indeed, a blade damage leads to additional acceleration contributions, namely $f_{as}^0$ and $f_{aa}^0$ for the linear acceleration and $\tau_a^0$ for the torque (thus, for the angular acceleration).

\subsection{Linear acceleration low frequency components}
Considering the matrix 
$\Xi_s=(\omegahat_{E,0}^0)^2 + \hat{\omegadot}_{E,0}^0$, it is possible to write $f_{as}^0 = \mbar\Xi_s (\sum_{i=1}^\na w_i l_i)$.
Thus, if $f_{as}^0$ is not identically null, then the \gls{UAV} $\Bcal$ is subject to an unbalanced chipped blade \gls{LOE}.
Therefore, the term $f_{as}^0$ is a symptom for detecting unbalanced chipped blade \gls{LOE}.
\begin{remark}
    Let $\Omega(\nu)= \Fcal[\omega_{E,0}^0]$ be the Fourier transform of the angular velocity and consider the transforms
	\begin{align}
		V_i(\nu) &= \Fcal[\omega_{E,0}^0\times(\omega_{E,0}^0 \times e_i)]\\
		S_i(\nu) &= V_i(\nu) - j2\pi \nu  (e_i\times \Omega(\nu)),
	\end{align}
	for $i=1,2$.
	Then, for null initial conditions the Fourier transform of $f_{as}^0$ is
	\begin{align}
		\Fcal[f_{as}^0] 
		&= \sum_{i=1}^{n_a} m_i(\inner{e_1,l_i^0} S_1(\nu) + \inner{e_2,l_i^0} S_2(\nu)).
	\end{align}
	Indeed, in the body frame the third component of $l_i^0$ is null, hence $l_i^0 = \inner{e_1,l_i^0}e_1 + \inner{e_2,l_i^0}e_2$.
	By direct calculation we have
	\begin{align*}
		\Fcal[f_{as}^0] 
		&= \sum_{i=1}^{n_a} m_i(\Fcal[\omega_{E,0}^0 \times(\omega_{E,0}^0\times l_i^0)] 
        + \Fcal[\omegadot_{E,0}^0\times l_i^0])\\
		&= \sum_{i=1}^{n_a}\sum_{k=1}^2 m_i\inner{e_k,l_i^0} \Fcal[\omega_{E,0}^0 \times(\omega_{E,0}^0\times e_k)] \\
        & -\sum_{i=1}^{n_a}m_i\inner{e_k,l_i^0} (e_k\times \Fcal[\omegadot_{E,0}^0])\\
		&= \sum_{i=1}^{n_a}\sum_{k=1}^2 
		m_i\inner{e_k,l_i^0}(V_k(\nu) - j2\pi\nu(e_k\times \Omega(\nu)))\\
		&= \sum_{i=1}^{n_a}\sum_{k=1}^2 
		m_i\inner{e_k,l_i^0} S_k(\nu).
	\end{align*}
	An unbalanced fault produces frequency components as a linear combination of $S_1(\nu)$ and $S_2(\nu)$, where the coefficients depend on the \gls{UAV} arms. 
\end{remark}

\begin{remark}
    The term $f_{as}^0$ depends on the faulty propellers arms, and therefore is a good candidate feature for the \gls{FDI}.
    However, $f_{as}^0$ is not directly affected by the rotors speed, and thus it can be considered as a low frequency component, with frequencies similar to those of the angular velocity.
    Each component is weighted by $m_i$ (and thus $m_i/m$ for the acceleration), which can easily be of order $10^{-2}$.
    This makes the \gls{FD} and \gls{FDI} using $f_{as}^0$ particularly challenging in a real world scenario.
\end{remark}

\subsection{Linear acceleration high frequency components}
    Let us define the matrices
    \begin{align}
        \Xi_{a,i} 
        &= 
        (\omegahat_{E,0}^0)^2 + \hat{\dot{\omega}}_{E,0}^0 + 2\thetadot_i \omegahat_{E,0}^0\ehat_3 +\thetaddot_i \ehat_3 - \thetadot_i^2  \mathcal{I}_3,
    \end{align}
    where $\mathcal{I}_3 $ is the identity matrix. For $i=1,\ldots,n_a$, it is possible to rewrite 
    \begin{align}
    	f_{aa}^0 &= \sum_{i=1}^\na m_ik_i\Xi_{a,i}p_i.
    \end{align}
    The force $f_{aa}^0$ is linear in $p_i$, which models the rotation of the blades.
    Depending on the \gls{UAV} maneuver, it is possible to greatly simplify such force contribution.
\begin{remark}
    If $f_{aa}^0$ is not identically null, then the \gls{UAV} $\Bcal$ is subject to an asymmetric blade fault.
    Therefore, the term $f_{aa}^0$ is a symptom for detecting asymmetrically chipped blades.
    The converse is however not true, since $\Xi_s$ depends on both $\omega_{E,0}^0$ and $\omegadot_{E,0}^0$, and thus it can point-wise vary its rank and null space.
\end{remark}
\begin{remark}
    Assume a propeller $\Bcal_i$ is subject to an asymmetric \gls{LOE}, while $\Bcal_j$ is healthy for each $j\neq i$.
    Forcing ideal hovering conditions, that is $\omega_{E,0}^0=\dot{\omega}_{E,0}^0=0$, $v_{E,0}^0=\vdot_{E,0}^0 =0$, $\thetadot_i=\thetadot_{i,hov}$ and $\thetaddot_i=0$, for some constant $\thetadot_{i,hov}\in\Rbb$, we have
    \begin{equation}\label{eq:puresinusoidal}
        f_{aa}^0 
        = 
        -k_i m_i\thetadot_{i,hov}^2 
        \col{\cos(\thetadot_{i,hov}\> t )\\\sin(\thetadot_{i,hov}\> t)\\0},
    \end{equation}
    and therefore a pure sinusoidal signals along the $x_0$ and $y_0$ axes.
    This confirms the approximation mostly used in the literature.
    Moreover, the first two components of the Fourier transform magnitude $|\Fcal[f_{aa}^0]|$ of $f_{aa}^0$ is a pure spike in the pulsation $\thetadot_{i,hov}$. 
    No vibrations are induced in the $z_0$ component. 
\end{remark}

During a general maneuver, the frequency components of $f_{aa}^0$ can be nontrivial.
Assume the rotor speed $\thetadot_i$ is constant over time, i.e., $\thetaddot_i= 0$.
Without loss of generality, also assume zero angle for each blade at the initial time, implying $\theta_i = \thetadot_i t$, and then $p_i = \cos(\thetadot_i t)e_1 +\sin(\thetadot_i t)e_2$.
Defining the signals
\begin{align}
    \gamma_{i,1}(t) &= m_ik_i\Xi_{a,i}e_1
    &
    \gamma_{i,2}(t) &= m_ik_i\Xi_{a,i}e_2,
\end{align}
it follows that
\begin{equation}
    \Fcal[f_{aa}^0(t)] = 
    \sum_{i=1}^\na 
    \left(
    \Fcal[\gamma_{i,1}(t) \cos(\thetadot_i t)] 
    + \Fcal[\gamma_{i,2}(t) \sin(\thetadot_i t)]\right).
\end{equation}
Therefore, the propeller rotation acts as an amplitude modulation on $\gamma_{i,1}(t)$ and $\gamma_{i,2}(t)$, and having $\theta_i$ as carrier pulsation.
Depending on which propeller $\Bcal_i$ is subject to a \gls{LOE}, the vibrations due to $f_{aa}^0$ may slide on the pulsation axis.
Therefore, a coordination between control actions and frequency domain residual generators can lead to fault isolation. 

\section{Fault Detection and Isolation}\label{sec:FDI}
In this section, we describe the proposed method to isolate \gls{LOE} faults due to blade chipping.
Firstly, we introduce the control allocation algorithm, as it plays a central role to perform active \gls{FDI}.
Then, the proposed \gls{FD} strategy, followed by the active \gls{FI} algorithm, are discussed.

\subsection{Control Allocation}\label{sec:allocation}
The control allocation is designed under the classic rigid body approximation as usual.
Even if the multirotor is a multibody system, the interaction between the propellers and the frame are by far negligible for the purpose of control.

Control allocation algorithms \citep{johansen2013control} are used to find the inputs $u \in \mathbb{R}^{n_a}$ such that
\begin{equation}\label{eq:controleffectiveness}
\tau = Bu,
\end{equation}
where $\tau \in \mathbb{R}^6$ is the wrench required by a control law, $B \in \mathbb{R}^{6 \times n_a}$ is the control effectiveness matrix, and $u = c_L\col{\thetadot_1^2 & \ldots & \thetadot_\na^2}$ denotes the motor lift forces.
\begin{remark}\label{rem:Allocation_noLOE}
    Blade chipping leads to \gls{LOE} and thus a loss of actuation wrench. 
    For the sake of simplicity, the input mapping \eqref{eq:controleffectiveness} is considered fault-free in performing control allocation, leading to a model mismatch that is negligible as long as the blade damage is small.
    However, in case of severe blade chipping \gls{LOE}, i.e., when an high percentage the blade is lost, the wrench mismatch can be significant, and a classical observer-based residual generator \cite{baldini2021disturbance} can be coupled with the proposed frequency domain active solution.
\end{remark}
Several methods can be used to solve \eqref{eq:controleffectiveness}, including the well-known Moore-Penrose pseudo-inverse \citep{baldini2020actuator}, which returns the solution $u$ with the minimum Euclidean norm.
However, to comply with constraints, such as saturation and rate limits, the control allocation problem is formulated as a Quadratic Programming (QP) problem \citep{baldini2020ifac}.

Let us define the decision variable
\begin{align}
x &= \begin{bmatrix} u \\ s \end{bmatrix} \in \mathbb{R}^{n_a + 1},
\end{align}
where $s\in\mathbb{R}$ is an artificial variable to relax equality constraints, thus preventing infeasibility.
Let us also define a quadratic cost function to be minimized
\begin{align}\label{eq:costfcn}
\min_{x} \quad J(x) = \frac{1}{2} x^\top H x + f^\top x,
\end{align}
with
\begin{align}\label{eq:QPparams}
H = 2 \begin{bmatrix}
\mathcal{I}_{n_a} & \mathbf{0} \\
\mathbf{0}^\top & \lambda_H
\end{bmatrix}, \quad
f = \begin{bmatrix}
\mathbf{0} \\
\lambda_f
\end{bmatrix},
\end{align}
where $\mathcal{I}_{\na}\in\mathbb{R}^{\na\times\na}$ is the identity matrix, $\mathbf{0}$ is a column vector of zeros with proper dimension, while $\lambda_H$ and $\lambda_f$ are scalar weights associated with the artificial variable $s$, which penalizes constraint violations in a manner similar to big-M methods \citep{nocedal1999numerical}.

Introducing the artificial variable $s$, the hard constraint \eqref{eq:controleffectiveness} is formulated as follows
\begin{align}\label{eq:ineq}
\begin{bmatrix}
-B & -\mathbf{1} \\
\hphantom{-}B & -\mathbf{1}
\end{bmatrix} x \leq
\begin{bmatrix}
-\tau \\
\tau
\end{bmatrix},
\end{align}
where $\mathbf{1}$ is a column vector of ones with proper dimension.
Essentially, \eqref{eq:ineq} is equivalent to $-\mathbf{1}s\leq Bu-\tau\leq\mathbf{1}s$ and $s\geq 0$ is implicitly required.
To enforce $s \approx 0$ whenever possible, so that \eqref{eq:controleffectiveness} holds,  $\lambda_H$ and $\lambda_f$ must be sufficiently large.

Control input and rate limits are imposed as box constraints:
\begin{align}
l_b \leq u \leq u_b,
\end{align}
where
\begin{align}\label{eq:saturations}
\begin{split}
l_b &= \max\left( u_{\text{prev}} - \delta, \mathbf{0}  \right)\\
u_b &= \min\left( u_{\text{prev}} + \delta, \mathbf{1} u_{\max} \right).
\end{split}
\end{align}
Both the saturation constraints in $[0,u_{\max}]$ and the rate limit $\delta$ are enforced in \eqref{eq:saturations}, where $u_{\text{prev}}$ is the previous value of $u$.
The problem \eqref{eq:costfcn}--\eqref{eq:saturations} is a conventional QP, therefore it can be solved by any suitable solver.

\subsection{Fault Detection}
To perform Fault Detection (\gls{FD}), we evaluate the individual terms of the \gls{DFT} of the acceleration signals, provided by the on-board accelerometers during regular flight.
The Goertzel algorithm is employed to compute the phase and amplitude in a range of frequencies including the motor speed at hovering.  
We then take the maximum amplitude and compare it to a fixed threshold $\rho_{\gls{FD}}$.
More precisely, for a closed interval $\Omega\subset \Rbb^+$ and for a time window length $T\in \Rbb^+$, consider the residual
\begin{equation}
    r_{\Omega,T}(t) = \max_{\omega\in \Omega} |\Fcal[\text{rect}(t-T,t)\vdot_{E,0}^0(t)]|,
\end{equation}
where $\Fcal[\cdot]$ denotes the Fourier transform and $\text{rect}(t-T,t)$ is the rectangle filtering the last $T$ s of signal. 
The residuals $r_{FD}$ and $r_{FDI}$ to perform the detection and isolation, respectively, are chosen as
\begin{align}
    r_{FD}(t) &= r_{\Omega_{FD},T}(t)\label{eq:rFD}\\
    r_{FDI}(t) &= r_{\Omega_{FDI},T}(t)\label{eq:rFDI}.
\end{align}
The frequency interval $\Omega_{FD}$, used for \gls{FD}, should be approximately centered around the rotational frequency at hover, and its width should be large enough to cover the relevant rotational frequencies encountered during flight.
In accordance to the vibration model, this choice allows the capture of vibrations associated with a faulty propeller while eliminating most frequency components related to maneuvers and filtering part of the sensor noise.
The frequency interval $\Omega_{FDI}$, used instead for \gls{FDI}, should lie outside the previous range and therefore it should not be excited during steady flight conditions.
By employing \gls{AFD}, one of the motors is forced to rotate within $\Omega_{FDI}$, thereby facilitating \gls{FDI}.
The numerical implementation of both $r_{FD}(t)$ and $r_{FDI}(t)$ are performed using the Goertzel algorithm. 
Since $\vdot_{E,0}^0(t)$ is unknown, the measured variable is used instead. 
When the threshold $\rho_{FD}$ is exceeded, the active \gls{FI} procedure, described in the remainder, is triggered.

\subsection{Active Fault Isolation}\label{sec:active}
Once a fault is detected, a \gls{FI} strategy is employed to identify the faulty propeller.  
The \gls{FI} strategy is active, meaning it involves a deliberate modification of the control input, and consists of three steps:
\begin{enumerate}[\arabic*)]
    \item Depending on the multirotor configuration, the \gls{UAV} could be stabilized in hovering before the active isolation is performed,
    \item the motor speeds are varied and the corresponding vibration is quantified,
    \item the motor showing the largest vibration is labeled as the faulty one.
\end{enumerate}

In the optional step 1), the \gls{UAV} could be stabilized in a hovering position for safety reasons, as well as to ease the \gls{FI}.

In step 2), we exploit actuator redundancy to vary the individual motor speeds, while preserving the nominal net wrench, acting on the QP parameters.
In fact, using the basic QP parameters shown in Section~\ref{sec:allocation}, the control effort would be equally weighted among the actuators.
Therefore, the motor speeds during hovering would be nearly identical, preventing the extraction of distinctive frequency signatures to isolate the faulty motor.

The objective of step 2) is to set $\thetadot_j = \thetadot_{des}$, or equivalently, to set a single motor lift force $u_j$ at a given $u_{des} = c_L \thetadot_{des}^2$.
This operation is repeated for each motor ($j=1,\ldots,\na$).
To do so, during the active \gls{FI} phase, we modify the QP parameters in \eqref{eq:QPparams} as follows
\begin{align}\label{eq:QPmod}
\begin{split}
    H_{j,j}&=\kappa\\
    f_j&=-\kappa u_{des}
\end{split}
\end{align}
with $0<\kappa<\lambda_H$.
In this way, the QP steers $u_j$ to $u_{des}$ whenever feasible.
In fact, rewriting \eqref{eq:costfcn}, we obtain
\begin{align}
\quad J(x) = \frac{\kappa}{2} u_j^2 - \kappa u_j u_{des} + \sum_{i\neq j}^{n_a} u_i^2 + \lambda_H s^2 + \lambda_f s.
\end{align}
Choosing a sufficiently large $\kappa$, $H_{j,j}$ and $f_j$ are large in magnitude, thus the solver prioritizes the minimization of $\frac{\kappa}{2} u_j^2 - \kappa u_j u_{des}$ whenever the problem is feasible (i.e., when $s=0$).
As $\frac{\kappa}{2} u_j^2 - \kappa u_j u_{des}$ is a parabola opening upwards, its unique global minimum is the only point where its gradient $\kappa u_j - \kappa  u_j u_{des}$ is zero, namely, $u_j=u_{des}$.

As $u_j=u_{des}$ holds, the motor lift force and speed of the $j$-th motor are known.
To perform \gls{FI}, we calculate the \gls{DFT} of the acceleration signals in a narrow range of frequencies around such known motor speed.
According to \eqref{eq:puresinusoidal}, we expect a sinusoidal signal centered at the motor speed, with a magnitude that increases along with the blade unbalance.
Therefore, repeating the test for each motor, the maximum amplitude corresponds to the faulty motor.

\begin{remark}
During the active isolation the condition $\thetadot_j = \thetadot_{des}$ is forced.
In order to improve the correct isolation rate, it is however required $\thetadot_i \neq \thetadot_{des}$ for $i\neq j$.
Thus, $\thetadot_{des}$ is chosen out from the range of speeds which is usually covered during the flight. 
In the extreme cases where this is not possible, additional constraints can be managed by the control allocation.
\end{remark}

\begin{remark}
    If the \gls{UAV} has sufficient input redundancy, an alternative solution is to set $\thetadot_{des}=0$ to completely remove the vibrations, instead of amplifying them during the active phase.
    In this case, the fault isolation residual should be constructed using a broad set of frequencies $\Omega_{FDI}$ and step~3) of the \gls{FI} strategy should be modified to operate using inverted logic.
    Depending on the multirotor configuration, this solution may result in temporary underactuation and difficulties in maintaining altitude and is therefore avoided.
\end{remark}

\section{Simulation results}\label{sec:simresults}
The method is tested using Matlab/Simulink on a octarotor (i.e., $n_a = 8$), whose parameters are reported in Table \ref{tab:parametersISA}.
The multirotor physical equations are simulated with a simulation step of $1$ $ms$, using the built-in Runge-Kutta solver.
Differently, the overall embedded code is implemented in discrete time through a backward Euler method with sample frequency of $200$ $Hz$.
This includes the control algorithm, the optimal allocation, and the fault detection and isolation residuals.

For the purpose of \gls{FD}, $\Omega_{FD} = [300,500]$ rad/s and $T=1$\>s are chosen in \eqref{eq:rFD}, so the Goertzel algorithm calculates the \gls{DFT} between 300 Hz and 500 Hz, with a step of 5 Hz.
When a fault is detected, with a threshold $\rho_{\gls{FD}}=0.005$, the active \gls{FI} strategy in Section~\ref{sec:active} is triggered, then the QP parameter modification in \eqref{eq:QPmod} is enabled in turn for each motor ($j=1,\ldots,8$) for $5$\>s each.
In particular, we employ $\lambda_H=100$, $\lambda_f=100$, and $\kappa=20$ as QP parameters.
MATLAB's \texttt{quadprog} function with the \texttt{active-set} algorithm is employed to solve QP problems.
In the meanwhile, as $\Omega_{FDI} = [500,540]$ and $T=1$\>s are selected, the Goertzel algorithm calculates the \gls{DFT} between 500 Hz and 540 Hz, with a step of 5 Hz to isolate the fault.
Finally, as considerable input redundancy is available for the octarotor under investigation, the optional step 1) in Section~\ref{sec:active} is avoided: we do not perform stabilization during \gls{AFD}, and the \gls{UAV} continues to track its trajectory.

The following aspects are taken into account to enhance the realism of the simulation. 
\begin{enumerate}
    \item The powertrain dynamics including the motors and the \glspl{ESC} are simulated as first order systems having time constant $\tau_{pt} = 0.1$ s. 
    \item 
    Each $i$th motor related to the propeller $\Bcal_i$ is unidirectional, and the related angular speed $\thetadot_i$ exhibits a saturation $-c_{w_i}\thetadot_i \in [0,620.97]$ rad/s. 
    Thus, the related motor upward lift force $f_i = ||f_{m_i}||$ constraint is $f_i\in [0,f_{max}]=[0,4.75]$ $N$, leading to a thrust to weight ratio equal to $2.5$. For each force $f_i$, we also simulate the PWM signal $PWM_i = f_i/f_{max}\in [0,1]$.
    \item 
    A time-varying external wrench, not available to either the control system or the \gls{FDI} module, is simulated to represent exogenous accelerations acting on the multirotor, such as those induced by wind disturbances, suspended payload dynamics, and similar effects.
    Accordingly, this term is mainly a low frequency acceleration and its Fourier transform is approximately zero for the pulsations near $\thetadot_{des}$.
    \item Additive sensor noise is simulated according to the datasheet of the MPU-9250 IMU \cite{invensense2014mpu}, which is commonly adopted by the commercial Cube autopilot (also known as Pixhawk 2 autopilot). Accelerometer and gyroscope noise is directly injected, while noise on the remaining state variables is assumed to be smaller due to Kalman filtering, as reported in Table \ref{tab:noiseMPU_9250_IMU}.
    \item That the IMU is placed at the \gls{COM} of $\Bcal_0$ (which is not the \gls{COM} of the \gls{UAV} in general), providing the measured variables $\vdot_{meas} = \vdot_{E,0}^0 + \nu_a$ and $\omega_{meas} = \omega_{E,0}^0 + \nu_\omega$, where $\nu_v,\nu_\omega$ denote white Gaussian noise.
\end{enumerate}
\begin{table}[tb]
	\centering
	\caption{Known hexarotor parameters \cite{niemiec2022multirotor}, known propeller parameters, and unknown parameters used for the simulation.}
	\label{tab:parametersISA}
	\begin{tabular}{l l l}
		\hline
		{Parameter} 					& {Value}		& {Meas. Unit}\\
		\hline
		Number of actuators ($n_a$)         & $8$                   & -\\
		Frame $\Bcal_0$ mass ($m_0$) 			& $1.55$ 				& kg \\
        Propeller $\Bcal_i$ mass ($m_i$) 			& $0.13$ 				& kg \\
		$\Bcal_0$ inertia along $x_B$ ($I_{0x}$) 		& $0.0266$ 				& kg m\textsuperscript{2} \\
		$\Bcal_0$ inertia along $y_B$ ($I_{0y}$)  		& $0.0266$ 				& kg m\textsuperscript{2} \\
		$\Bcal_0$ inertia along $z_B$ ($I_{0z}$)  		& $0.0498$ 				& kg m\textsuperscript{2} \\
		Gravitational acceleration ($g$)	& $9.81$ 				& m/s\textsuperscript{2} \\
		Arm length ($l$)					& $0.275$ 				& m \\
		Propeller $\Bcal_i$ radius ($r$) 			& $0.12$ 				& m \\
        Propeller $\Bcal_i$ width ($a$) 			& $0.025$ 				& m \\
        \hline
		Lift coefficient ($c_L$)        & $1.23\cdot 10^{-5}$       & N s\textsuperscript{2}\\
		Drag coefficient ($c_D$)        & $1.10\cdot 10^{-7}$       & N m s\textsuperscript{2}\\
		\hline
		Friction coefficient ($k_t$)        & $3.2\> 10^{-2}$                   & s\textsuperscript{-1}\\
		Friction coefficient ($k_r$)        & $5.57\> 10^{-4}$                   & s\textsuperscript{-1}\\
		\hline
	\end{tabular} 
\end{table}
\begin{table}[tb]
	\centering
	\caption{Additive output white gaussian noise.}
	\label{tab:noiseMPU_9250_IMU}
	\begin{tabular}{ll}
		\hline
		{Affected variables}\hspace{10mm} & {Standard deviation}\\
		\hline
		Components of $\vdot_{E,0}^0$ &  $0.0785$\\
		Components of $v_{E,0}^0$ &  $0.0392$\\
		Components of $p_{E,0}^E$ &  $0.0196$\\
		Components of $\omega_{E,0}^0$ & $0.0055$\\
		Roll, pitch, yaw angles & $0.0028$\\
		\hline
	\end{tabular}   
\end{table}

\subsection{Motors and \glspl{ESC}}
In the simulations, each pair composed by an \gls{ESC} and a motor is modeled as a first order system 
\begin{align}
	\label{eq:ESC}
	\thetaddot_i &= -\lambda_i (\thetadot_i - 	\thetadot_{i,r}) & &i=1,\ldots,\na,
\end{align}
where $\lambda_i\in\Rbb$, and $\thetadot_{i,r}$ denotes the desired angular speed of the $i$th propeller.
Note that the \gls{ESC} equation \eqref{eq:ESC} does not provide the convergence of $|\thetadot_{i}-\thetadot_{i,r}|\to 0$ unless $\thetaddot_{i,r}=0$, which models some possibly open loop components (e.g., the effect of the inertia tensors contribution $I_i$ of each propeller). 
The references $\thetadot_{1,r},\ldots,\thetadot_{\na,r}$ represent the control inputs for the system.

\subsection{Time Domain Plots}
Using a double loop feedback linearization coupled with PID controllers, the octarotor travels an ascending helicoid trajectory.
An asymmetric \gls{LOE} is introduced at $t=10$ s on the propeller $\Bcal_2$. 
The chipping, affecting a single blade, is abrupt and equal to $10\%$ of the propeller's blade radius. 
In the following time domain plots, the time interval $I_{FDI} = [10.22,31.22]$ s is highlighted with a red background. 
$I_{FDI}$ represents the overall time required for the fault detection and the active isolation.

\begin{figure}[t]
	\centering
	\includegraphics[width=1\linewidth]{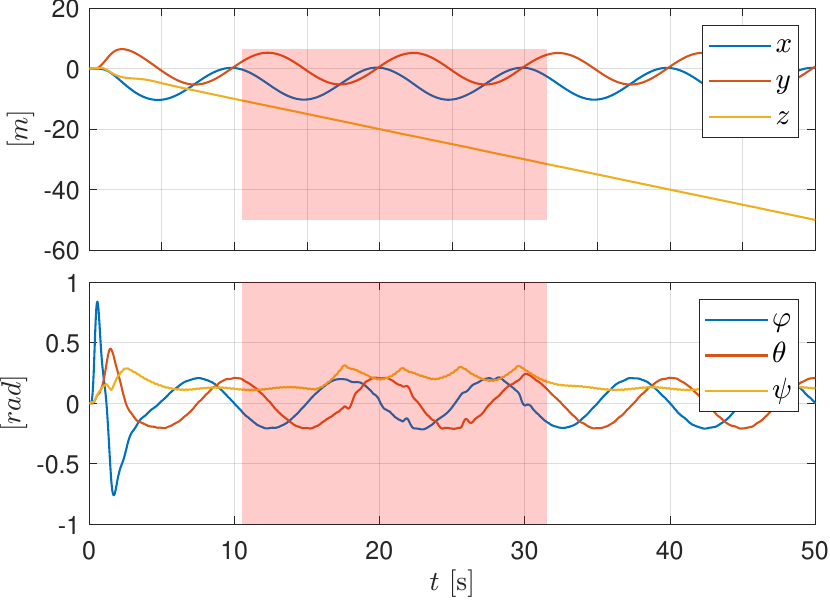}
	\caption{Pose of the multirotor. } 
	\label{fig:pose}
\end{figure}

\begin{figure}[t]
	\centering
	\includegraphics[width=1\linewidth]{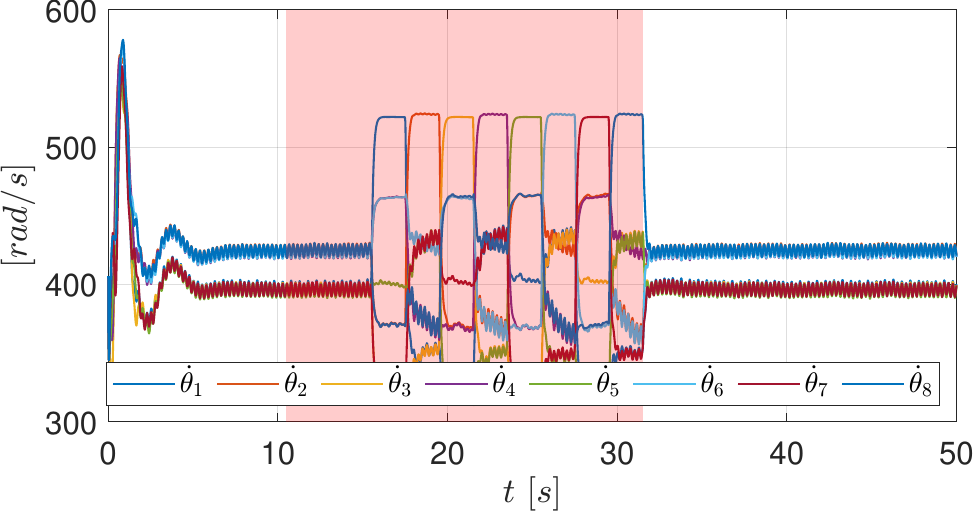}
	\caption{Octarotor motors angular speeds. During the active \gls{FDI} the propellers speeds are allocated accordingly with the optimal control allocation previously described.}
	\label{fig:MotorsSpeeds}
\end{figure}

\begin{figure}[t]
	\centering
	\includegraphics[width=1\linewidth]{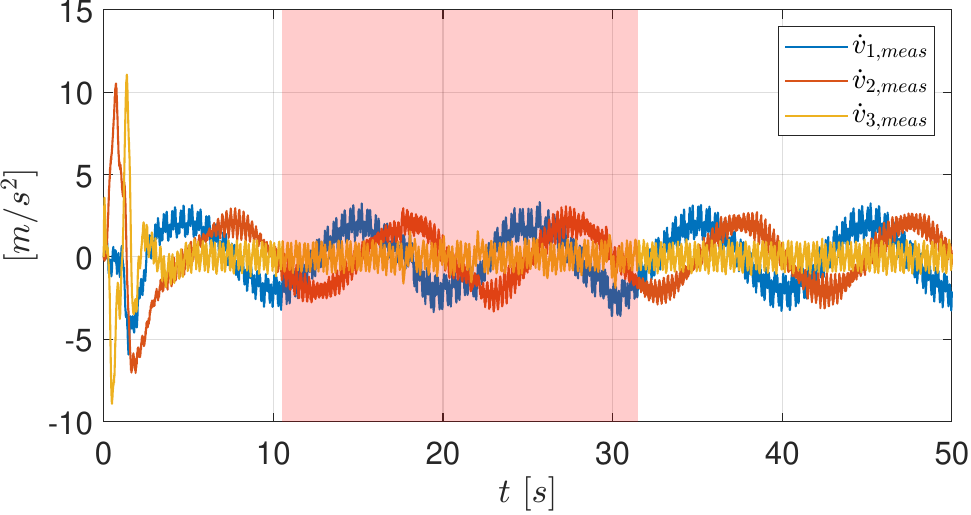}
	\caption{Measured linear body acceleration components.}
	\label{fig:LinearAccelerations}
\end{figure}

\begin{figure}[t]
	\centering
	\includegraphics[width=1\linewidth]{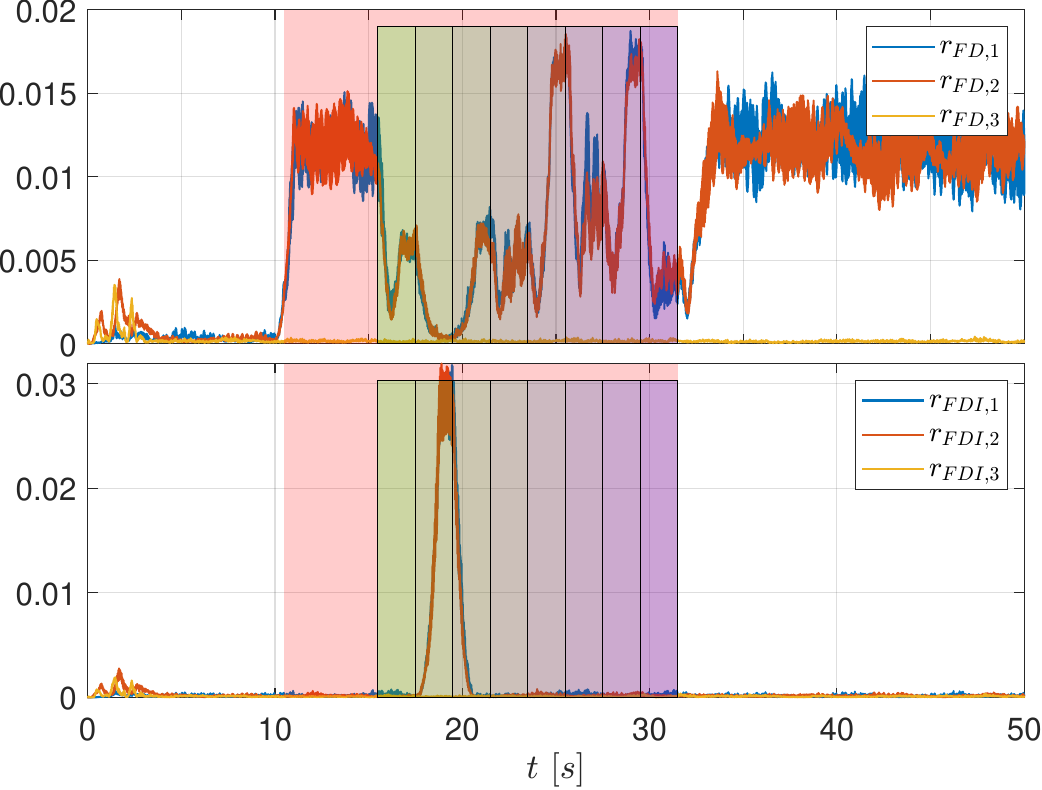}
	\caption{\gls{FD} residual and \gls{FDI} residual. The \gls{FD} residual triggers the active \gls{FDI} stage. In this scenario the actual faulty motor is the second one, and the \gls{FDI} residual triggers in the second stage of the active isolation.}
	\label{fig:Residuals}
\end{figure}

The pose of the multirotor is depicted in Fig.~\ref{fig:pose}.
For $t\in I_{FDI}$, the octarotor smoothly continues to follow the trajectory, shifting the active isolation burden to the allocation. 
Once the \gls{FD} triggers, the active isolation starts, and the motors angular speeds are allocated as reported in Fig.~\ref{fig:MotorsSpeeds}.
During each stage of the \gls{FDI}, the rotor's speed $\thetadot_j$ of the candidate $j$th motor is set approximately to $\thetadot_{des}$, separating it from the remaining rotors' speeds.
The measured linear body accelerations $\vdot_{meas}$ are represented in Fig.~\ref{fig:LinearAccelerations}.
The asymmetric injected fault at $t = 10$ s leads to additional high frequency terms, and the detection can be performed in the frequency domain, here investigated using the \gls{DFT}.
Finally, both the detection residual $r_{FD}$ and the isolation residual $r_{FDI}$ are plotted in Fig.~\ref{fig:Residuals}.
To better clarify the active strategy, the background color is set according to the motor under investigation. 
Once $r_{FD}$ has been triggered, the active isolation starts and the second \gls{DFT} algorithm around the prescribed angular speed pulsation $\thetadot_{des}$ provides $r_{FDI}$.
In this simulation the faulty propeller is $\Bcal_2$, and therefore $r_{FDI}$ clearly exceeds the threshold for $t\in I_{FDI,2}$ only.

\subsection{Frequency Domain Plots}
\begin{figure*}
\centering
\includegraphics[width=1\linewidth]{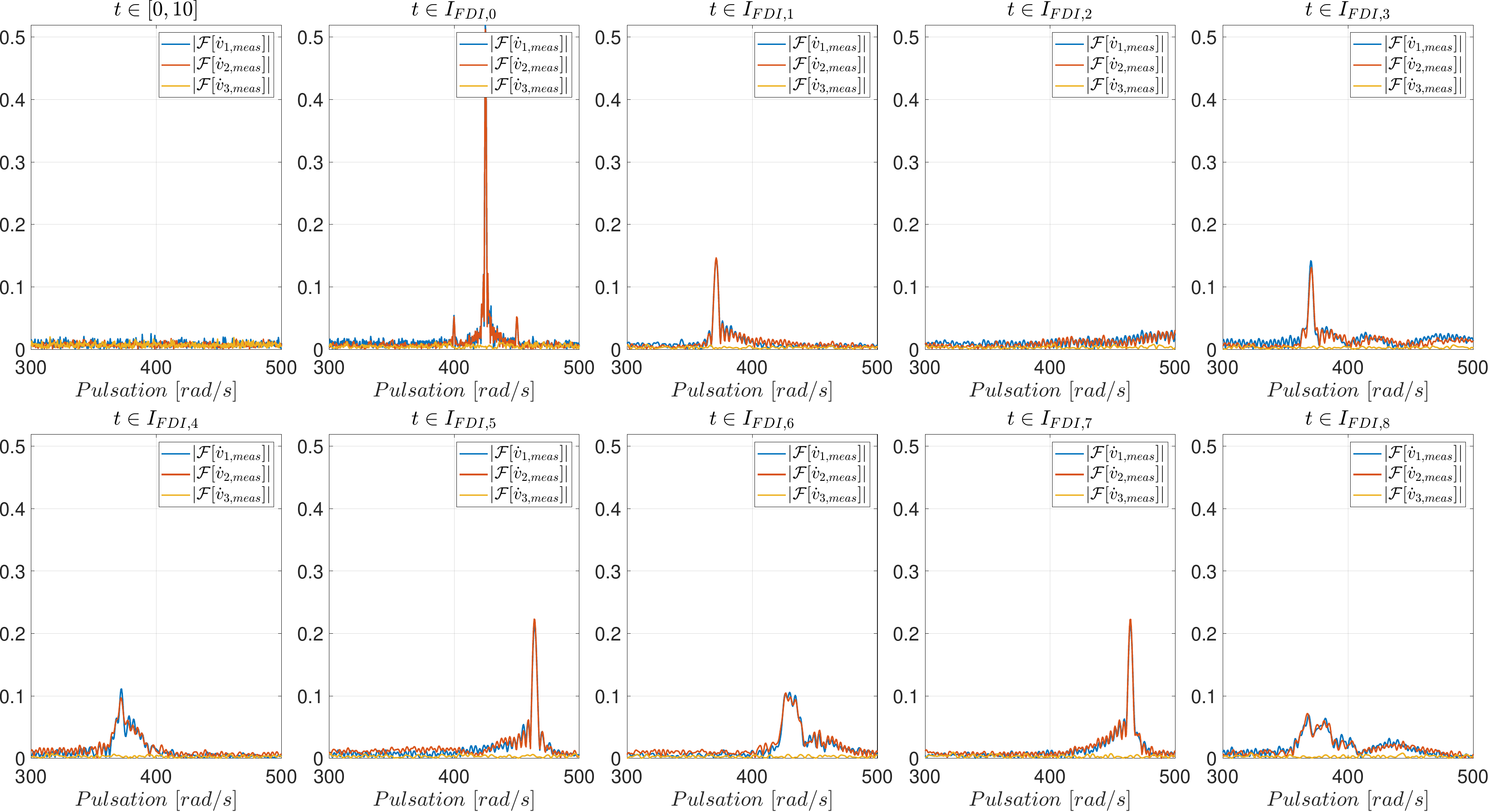}
	\caption{Fourier magnitude of measured acceleration before the fault, after the fault, and during each active isolation stage, over the frequency range used for fault detection.
    }
	\label{fig:FT_FD}
\end{figure*}

\begin{figure*}
\centering
\includegraphics[width=1\linewidth]{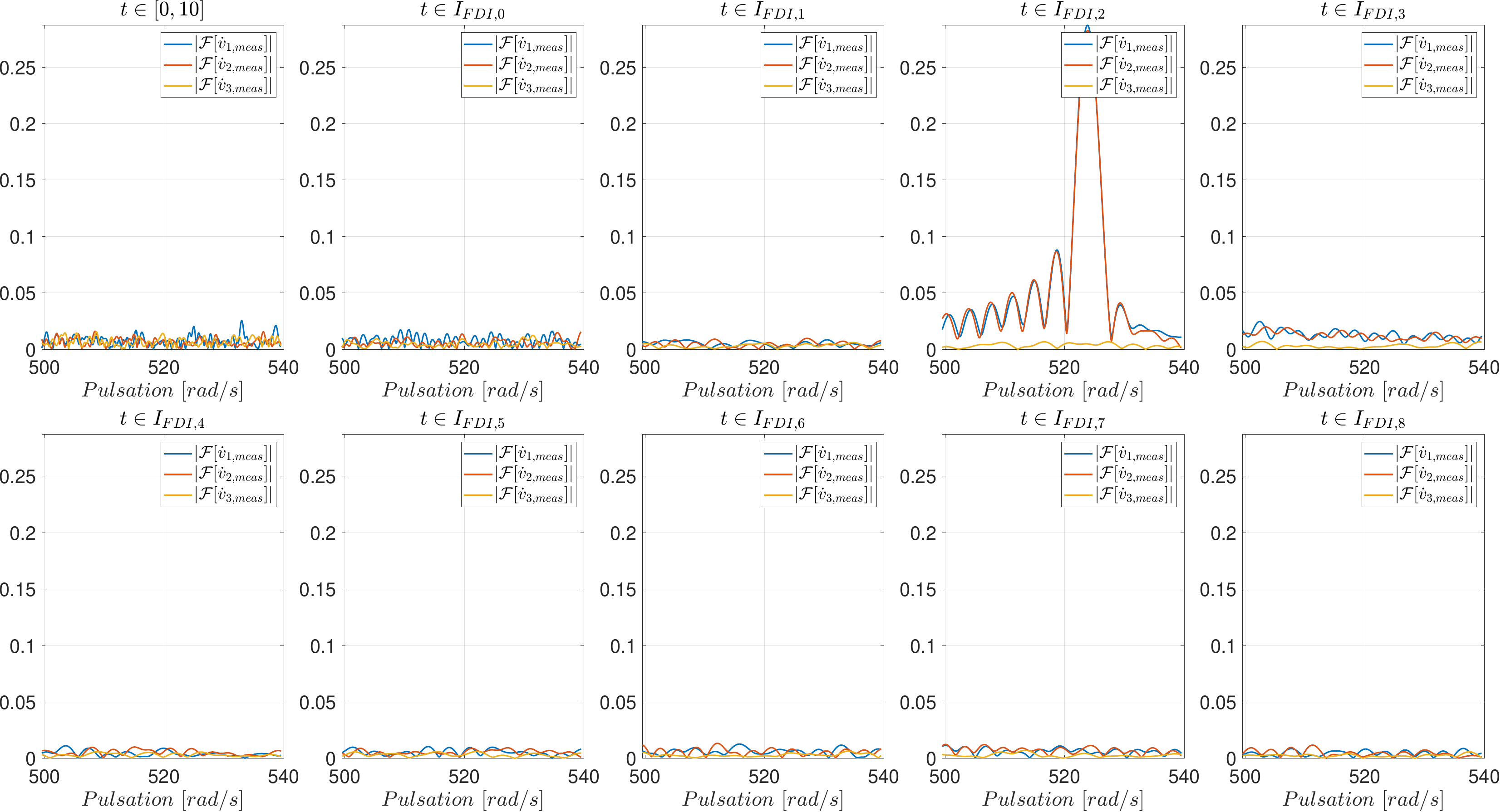}
	\caption{Fourier magnitude of measured acceleration before the fault, after the fault, and during each active isolation stage, over the frequency range used for active fault isolation.
    }
	\label{fig:FT_FDI}
\end{figure*}

Fault detection and active isolation rely on the coordinated excitation and analysis of the frequency components inherent to the multirotor dynamics.
The magnitude of the Fourier transforms of $\vdot$ is then reported in Fig.~\ref{fig:FT_FD}.
Although the body linear acceleration $\vdot$ has three components, the last one is almost negligible for detection purposes.
In fact, the vibrations propagate mainly along the $x_0$ and $y_0$ axes.
Since the spectrum varies w.r.t. the rotors' angular speeds $\thetadot_i$, a total of $10$ time windows are reported separately.
The first time window $[0,10]$ s represents the fault-free case.
The frequency components in the pulsation range $\Omega_{FD}=[300, 500]$ rad/s are due to the noise and the regular \gls{UAV} motion.
Clearly, the average propellers' angular speeds obtained in a near hovering motion are within the range $\Omega_{FD}$. 
The second plot in Fig.~\ref{fig:FT_FD} is for $t\in I_{FDI,0}$, representing the time when the fault has been injected. 
Since the amplitude is highly increased due to the vibration, the \gls{FD} residual triggers. 
All the remaining plots are relative to the time windows $I_{FDI,i}$, for $i=1,\ldots,8$, representing the active isolation phase for which the allocation goal is $\thetadot_i=\thetadot_{des}$, in addition to the control commands.
Spurious pulsations in each time window can appear, and they are not easily related to the faulty propeller $\Bcal_2$. 
Therefore the pulsation range $\Omega_{FDI}=[500,540]$ rad/s is considered. Fig.~\ref{fig:FT_FDI} reports the magnitude of the Fourier transform of $\vdot$ for the same time windows, but in the pulsation range $\Omega_{FDI}=[500,540]$ rad/s.
The condition $\thetadot_{des}=520$ rad/s is imposed by the active isolation strategy, and the isolation residual $r_{FDI}$ is indeed calculated centered on this pulsation range.
In the time interval $t\in I_{FDI,2}$, the magnitude increases significantly, proving that fault isolation is feasible.

\subsection{Statistical Analysis}
Fault detection and isolation can present additional challenges depending on the boundary conditions, such as the fault magnitude, the signal to noise ratio, the damping of mechanical vibrations, and the desired trajectory. 
To evaluate the robustness and performance of the proposed method, a total of $9600$ simulations were conducted under varying conditions. These conditions include $4$ tracking references (straight lines, ascending helicoids, ascending $8$-shapes, squares), $5$ \gls{LOE} magnitudes affecting one blade (ranging from $0\%$ to $20\%$ unilateral chipping, where $0\%$ means no chipping), $3$ possible vibration damping factors $d$ (defined as the ratio of the measured acceleration amplitude to the actual acceleration amplitude), and $20$ tentative thresholds $\rho_{FD}$ (ranging from $0.5\cdot 10^{-3}$ to $10^{-2}$). 
For each combination of the aforementioned parameters, each one of the $n_a=8$ propellers has been simulated under the same conditions, i.e., $1200$ simulations per motor, where $240$ are fault-free and $960$ with chipping appearing and $t=10$ s.
All the remaining boundary conditions (controller, external wrench, etc.) are the same as previously described in Section~\ref{sec:simresults}. 

\begin{figure*}[t]
\centering
\begin{subcaptionbox}[0.32\textwidth]{$d=5\%$.\label{fig:Damping950}}
    {\includegraphics[width=0.32\textwidth]{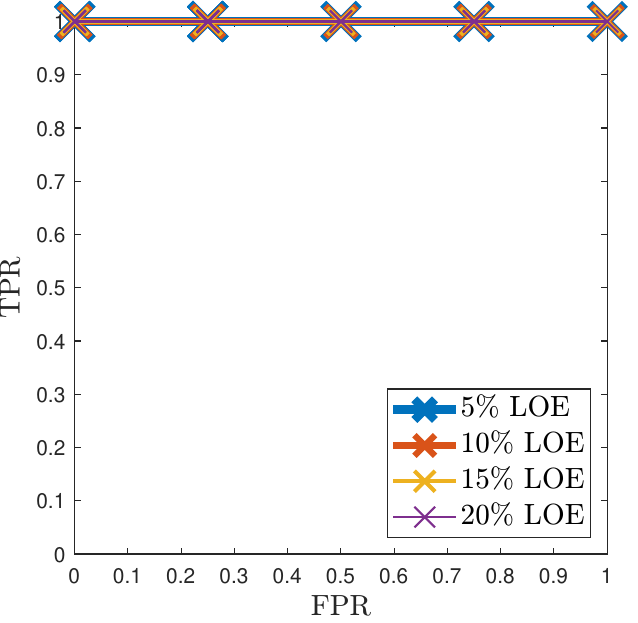}}
\end{subcaptionbox}
\hfill
\begin{subcaptionbox}[0.32\textwidth]{$d=3\%$.\label{fig:Damping970}}
    {\includegraphics[width=0.32\textwidth]{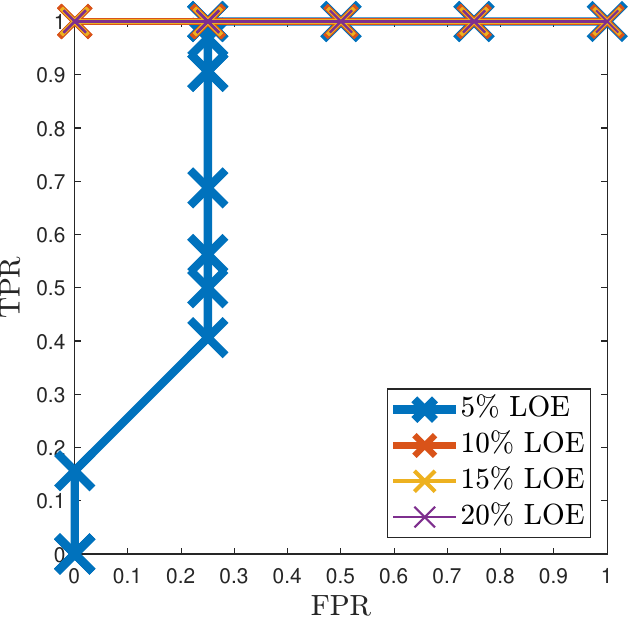}}
\end{subcaptionbox}
\hfill
\begin{subcaptionbox}[0.32\textwidth]{$d=1\%$.\label{fig:Damping990}}
    {\includegraphics[width=0.32\textwidth]{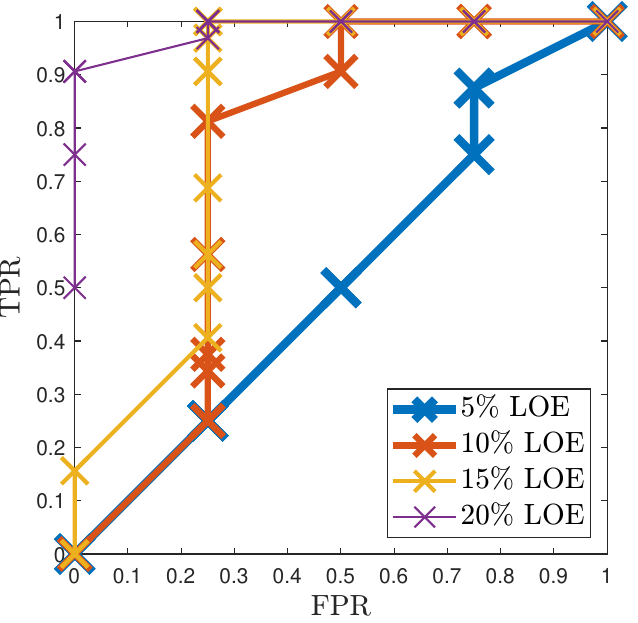}}
\end{subcaptionbox}
\caption{ROC curves for different damping factors $d$.}
\label{fig:Damping}
\end{figure*}

Fig.~\ref{fig:Damping} reports the ROC curves to determine the optimal threshold $\rho_{FD}$ under different damping factors $d$, where $d=5\%$ means that the $95\%$ of vibration amplitude is absorbed by the non-rigid structure and the \gls{IMU}, thus the measured vibration amplitude is only $5\%$ of the one calculated using the rigid-multibody equations.
The \gls{TPR} is defined as the ratio between correct fault isolation occurrences (the algorithm isolated a fault and the fault was actually present) and the total number of flights affected by a fault.
The \gls{FPR}, instead, is defined as the ratio between wrong fault isolation occurrences (the algorithm isolated a fault while none were present) and the total number of flights without faults.
Each point of the ROC curve represents a tentative threshold value $\rho_{FD}$.
The curves are parameterized by the fault amplitude.
Considering the smallest attenuation $d=5\%$, we obtain in Fig.~\ref{fig:Damping950} a ROC curve representing a perfect classifier for each fault magnitude.
In other words, there exists a threshold $\rho_{FD}\in[0.0080, 0.01]$ such that the accuracy is maximal, i.e., $\text{TPR}=1$ and $\text{FPR}=0$.

Clearly, as the attenuation increases, \gls{FDI} is more challenging due to the worse signal-to-noise ratio.
Fig.~\ref{fig:Damping970} shows that, in case of the smallest \gls{LOE}, there is no threshold $\rho_{FD}$ achieving perfect accuracy when the damping $d$ is $3\%$.
In this case, we still obtain no false positives using $\rho_{FD}=0.0080$, but the \gls{TPR} drops to $15.62\%$ when the \gls{LOE} is small ($5\%$) as the \gls{FD} residual does not overcome the threshold.
For larger \glspl{LOE}, instead, $\rho_{FD}=0.0080$ still achieves perfect isolation.
On the other hand, a smaller $\rho_{FD}=0.0040$ maximizes the \gls{TPR} to $100\%$, but a significant $25\%$ \gls{FPR} arises.
Practically, choosing a large $\rho_{FD}=0.0080$ is preferable, thus avoiding false positives and accepting the fact that minor faults can go unnoticed.

A $1\%$ damping makes the proposed algorithm less effective, as shown in Fig.~\ref{fig:Damping990}.
If avoiding false positives is prioritized ($\rho_{FD}=0.0080$), even some of the largest $20\%$ \gls{LOE} faults can go unnoticed ($90.62\%$ \gls{TPR}).
Moreover, for a $5\%$ \gls{LOE}, the method behaves like a random classifier, as the ROC mostly coincides with the line of no-discrimination.

\begin{figure*}[t]
\centering
\begin{subcaptionbox}[0.32\textwidth]{$d=5\%$.\label{fig:FDI_ConfusionMatrix1}}
    {\includegraphics[width=0.32\textwidth]{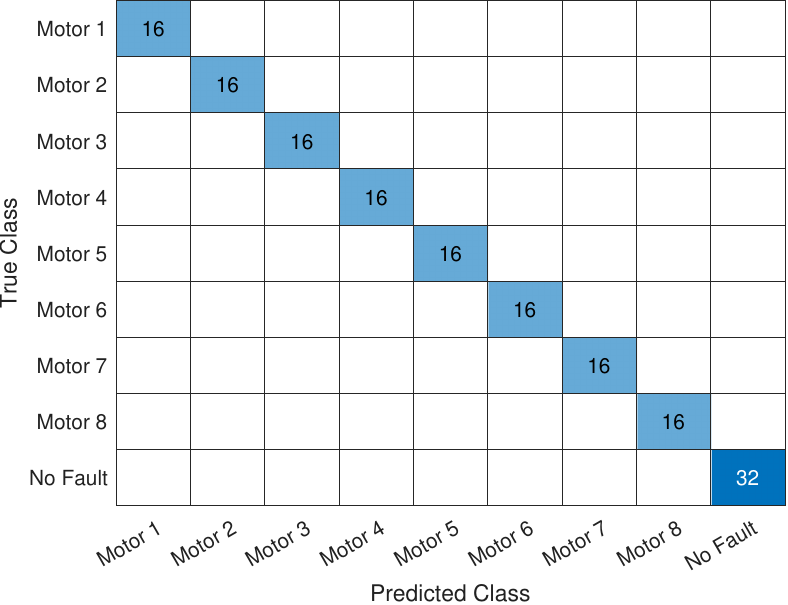}}
\end{subcaptionbox}
\hfill
\begin{subcaptionbox}[0.32\textwidth]{$d=3\%$.\label{fig:FDI_ConfusionMatrix2}}
    {\includegraphics[width=0.32\textwidth]{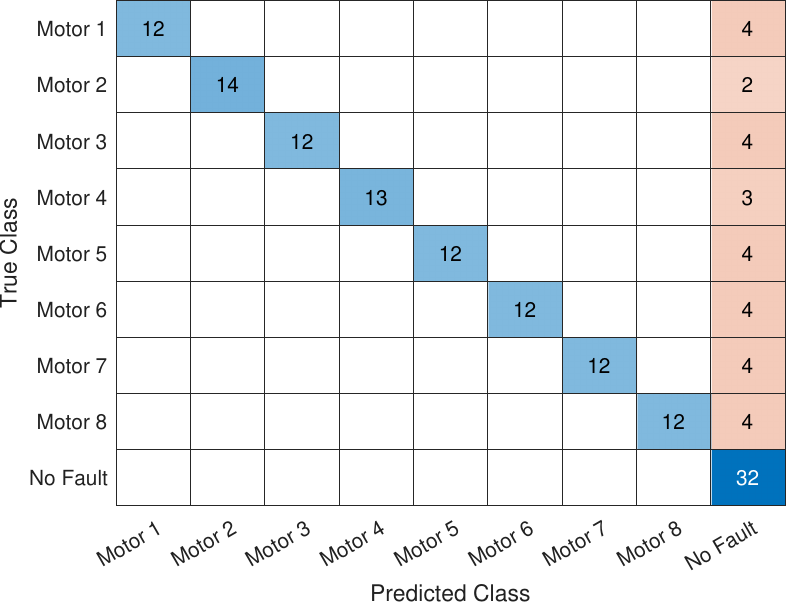}}
\end{subcaptionbox}
\hfill
\begin{subcaptionbox}[0.32\textwidth]{$d=1\%$.\label{fig:FDI_ConfusionMatrix3}}
    {\includegraphics[width=0.32\textwidth]{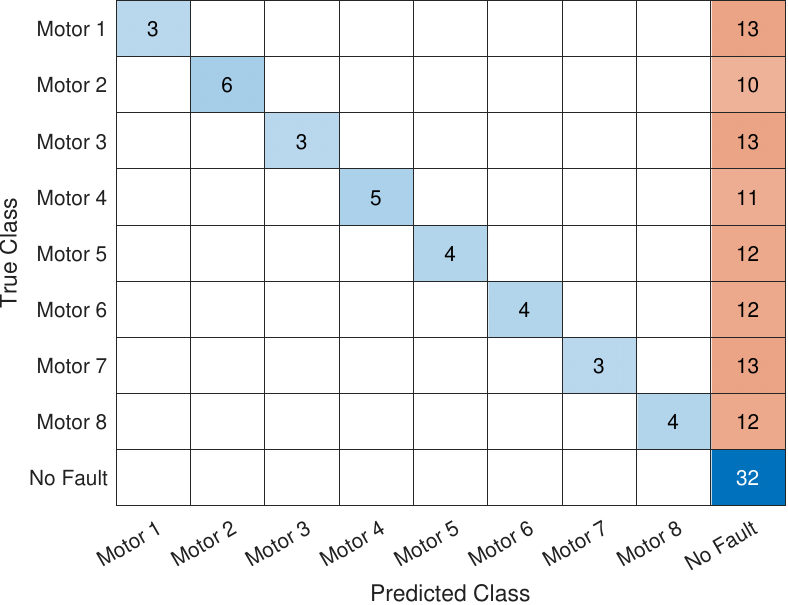}}
\end{subcaptionbox}
\caption{Confusion matrix choosing $\rho_{FD}=0.0080$.}
\label{fig:FDI_ConfusionMatrix}
\end{figure*}
Fig.~\ref{fig:FDI_ConfusionMatrix} reports the confusion matrix regarding fault isolation, choosing $\rho_{FD}=0.0080$ to avoid false positives according to Fig.~\ref{fig:Damping}.
False negatives are distributed among all the propellers with no pattern.
The false negative rates are $0\%$, $16.875\%$, and $58.75\%$, respectively with damping $5\%$, $3\%$, and $1\%$.
In Fig.~\ref{fig:FDI_ConfusionMatrix}, where every misclassification is a false negative, i.e., missed fault detection, while no confusion between faulty actuators appears in the isolation phase.

\begin{figure}[t]
	\centering
	\includegraphics[width=1\linewidth]{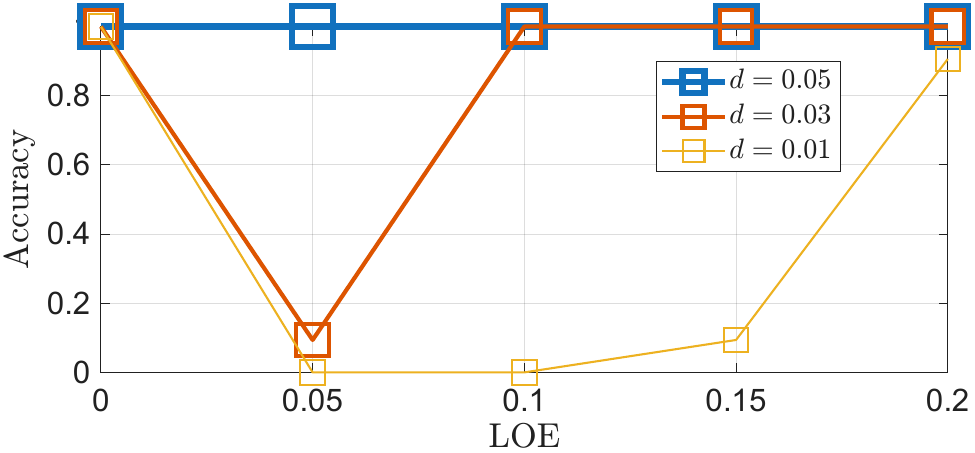}
	\caption{
        Correct \gls{FDI} ratio w.r.t. the LOE severity and the damping $d$, choosing $\rho_{FD}=0.0080$.}
	\label{fig:FDI_Percentage_Damping}
\end{figure}

The overall accuracy of the solution choosing $\rho_{FD}=0.0080$ is resumed in Fig.~\ref{fig:FDI_Percentage_Damping} for different \gls{LOE} severity and damping factors.
The accuracy is higher as the fault magnitude increases, as expected, due to the increasing vibrations.
Also, high attenuation affects negatively the accuracy: a damping $d=5\%$ allows for perfect classification, $d=3\%$ makes possible to classify exactly \glspl{LOE} starting from $10\%$, while in the worst case ($d=1\%$) no \glspl{LOE} strictly less than $20\%$ can be isolated. 

\section{Conclusion}\label{sec:conclusion}
This work presented a comprehensive analysis of how blade damage affects the dynamics of multirotor vehicles, and introduced a model-based active FDI strategy that leverages vibration data from the on-board IMU.
Starting from a white-box model capable of capturing both the loss of aerodynamic efficiency and the characteristic vibrations induced by propeller faults, a sensor-minimal active diagnostic method has been designed, making fault isolation possible by deliberately injecting controlled perturbations into the system.

The proposed FDI approach was evaluated in simulation on an octarotor platform using a dataset of 9600 scenarios. Both time and frequency domain features were extracted and analyzed, showing that vibration signatures provide reliable indicators of blade faults.

The method demonstrated high accuracy under nominal damping conditions; however, its performance degrades in the presence of strong attenuation, particularly when small-magnitude faults are considered, due to the attenuation of fault-induced vibrations.
This degradation is primarily reflected in an increased false-negative rate, as the vibration signatures become less distinguishable from nominal system behavior.
The confusion matrices indicate that fault detection represents the most critical step; conversely, once the presence of a fault has been identified, the active isolation stage proves robust, making fine-tuning of $\thetadot_{des}$ unnecessary.
In future work, we aim to address this limitation by incorporating an active fault detection strategy to facilitate detection under poor signal-to-noise ratio conditions.
When applied at regular time intervals, this approach may enhance the fault detection rate while limiting the increase in energy consumption associated with energy-suboptimal thrust allocation.

\FloatBarrier

\bibliographystyle{elsarticle-harv} 
\bibliography{bibliografia}
\balance

\end{document}